\documentclass[doublecol]{epl2} 

\usepackage{psfrag}
\usepackage{graphicx}
\usepackage{color}

\usepackage[latin1]{inputenc}
\usepackage{graphicx}
\usepackage{amsmath}
\usepackage{amssymb}
\usepackage{relsize}

\newcommand{\beq}{\begin{equation}}
\newcommand{\eeq}{\end{equation}}

\newcommand{\Wc} {W_{\rm{chem}}}

\newcommand{\nn} {\nonumber}

\newcommand{\Fm} {F^*}
\newcommand{\Fst} {F^{\rm{st}}}

\newcommand{\Pc} {{\dot W}_{\rm{chem}}}

\begin{document}

	\title{Efficiency of molecular motors at maximum power}
\author{Tim Schmiedl and Udo Seifert}
\institute{                    
  {II.} Institut f\"ur Theoretische Physik, Universit\"at Stuttgart,
  70550 Stuttgart, Germany
}

\begin{abstract}
{Molecular motors transduce chemical energy obtained from hydrolizing
$ATP$ into mechanical work exerted against an external force. 
We calculate their efficiency at maximum power output for two
simple generic models and show that the qualitative behaviour depends crucially
on the position of the transition state  or, equivalently, on the load distribution factor. 
Specifically, we find a transition state near the initial state 
(sometimes characterized as a ``power stroke'')
to be most favorable with respect to both high power output and high efficiency at maximum
power. In this regime, driving the motor further out of equilibrium
by applying higher chemical potential differences can even,
counter-intuitively, increase the efficiency.}
\end{abstract}

 {Nonequilibrium and irreversible thermodynamics}
 {Fluctuation phenomena, random processes, noise, and Brownian motion}
 {Subcellular structure and processes}

\def\la{\lambda(\tau)}
\def\ls{\lambda^*(\tau)}
\def\li{\lambda_i}
\def\wmus{W[\la]}

\maketitle

\section{Introduction}
Molecular motors are essential for directed transport within the cell \cite{howard}. 
They typically operate under nonequilibrium conditions due to the unbalanced chemical potentials of molecules like $ATP$ or $ADP$ involved in the chemical reactions accompanying the motor steps. In contrast to macroscopic engines, fluctuation effects are important thus allowing for backward steps even in directed motion. The stochastic dynamics of these motors under an applied load force can be probed experimentally by single molecule assays (see, e. g., for kinesin \cite{cart05}, myosin \cite{meht99, clem05, gebh06} or ATPase \cite{yasu98}). Generically, such biomotors are modelled either in terms of continuous ``flashing ratchets'' \cite{astu94, juel97, astu02, reim02a} or by a (chemical) master equation on a discrete state space \cite{fish99, lipo00, fish01, kolo03 , seif05, liep07a, lau07a}.

For macroscopic engines working between two heat baths at temperatures $T_2 > T_1$, efficiency is bounded by the Carnot limit $\eta_C = 1- T_1 / T_2$. Since this limit can only be achieved by driving the engine infinitesimally slowly, thus leading to an infinitesimally small power output, it is arguably more meaningful to characterize engines by their efficiency at maximum power \cite{curz75, hoff03}. 
This quantity has been studied for more than 30 years under the label of ``finite-time thermodynamics'' \cite{curz75, band82, andr84, leff87, hoff03}. Recently, this concept has been transferred to microscopic (Brownian) heat engines in a variety of different model systems \cite{vela01, gome06, schm08}.

In contrast to heat engines, biomotors are driven by chemical potential differences. The efficiency of such motors is bounded by $\eta_{\rm{max}} = 1$ \cite{parm99}. This bound can only be reached in an equilibrium situation corresponding to a vanishing power output of the motor. In analogy with heat engines, we here propose to investigate such motors under the condition of maximum power output.

We start with a simple model system for a chemically driven biomotor \cite{fish99} and show that the qualitative results also apply to a more realistic motor model involving a second cycle.  In this second model, a futile cycle leads to $ATP$ consumption even at the stall force. Mechanical and chemical cycles are no longer tightly coupled. In both cases, the efficiency at maximum power crucially depends on the position of the transition state or, equivalently, on the load distribution factor. In fact, a transition state near the initial position is most favorable with respect to a maximal motor power output. For the efficiency at maximum power, we obtain two counter-intuitive results : (i) it increases when the transition state position is changed in such a way that the power output rises and (ii) it can increase when the system is driven further out of equilibrium by a higher chemical potential difference.

\section{Model I}
\begin {figure}
\begin{center}
\includegraphics[width = 0.49 \textwidth]{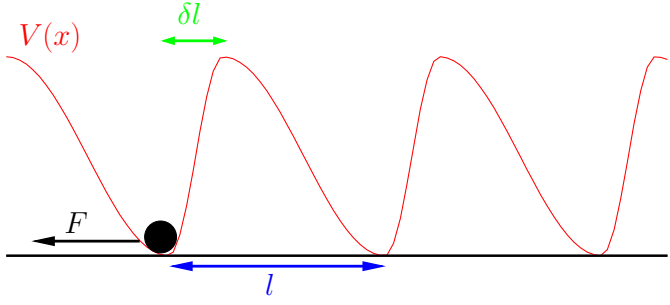}
\caption{(Color online) Scheme of the free energy landscape $V(x)$ of a molecular motor with step size $l$. The transition state position $\delta$ determines the force dependence of the transition rates. 
\label{fig0}}
\end{center}
\end {figure}

We first consider a linear molecular motor with equivalent discrete states (sites) $X_n$ ($n=0, \pm 1, \pm 2,\dots$) with distance $l$ between the sites and next-neighbour transitions between these states subject to a force $F$ in backward direction, see Fig \ref{fig0}. Forward reactions are assumed to be driven by $ATP$ molecules with chemical potential $\mu_{\rm{ATP}}$ and backward transitions by $ADP$ and $P$ molecules with chemical potentials  $\mu_{\rm{ADP}}$ and $\mu_{\rm{P}}$, respectively, 
\beq
ATP + X_n \mathop{\rightleftharpoons}_{w^-}^{w^+} X_{n+1} + ADP + P.
\eeq 
If the dilution of all involved species is high, we can assume mass action law kinetics for the rate constants. Additionally, we assume the usual force dependence of rate constants \cite{fish99} such that the transition rates for forward and backward steps are given by
\begin{eqnarray}
w^+ &=& c_{\rm{ATP}} k^+ e^{-\beta \delta F l} \\
w^- &=& c_{\rm{ADP}} c_{\rm{P}} k^- e^{\beta (1-\delta) F l},
\end{eqnarray}
respectively. Here, $c_i$ are the (dimensionless) concentrations of $i$ molecules ($i = \{ATP, ADP, P\}$ ) and $\beta \equiv 1 / k_{\rm{B}} T$ with Boltzmann's constant $k_{\rm{B}}$. The bare reaction rates $k^+$, $k^-$ are concentration independent. The load distribution factor $0 \leq \delta \leq 1$ characterizes the location of the transition state, see Fig. \ref{fig0}. It can vary between the extreme cases $\delta = 0$ (sometimes characterized as a ``power stroke'' \cite{howa06} ) and $\delta = 1$, where forward or backward rate constants, respectively, no longer depend on the force. The chemical potential of the involved molecules is
\begin{eqnarray}
\mu_{i} = \mu_{i}^0 + k_{\rm{B}} T \ln c_{i} , 
\end{eqnarray}
with a reference value $ \mu_{i}^0$. Thermodynamic consistency requires 
\beq
w^+ / w^- = e^{( \Delta \mu - F l) / (k_{\rm{B}} T)},
\eeq
where $\Delta \mu \equiv \mu_{\rm{ATP}} - \mu_{\rm{ADP}} - \mu_{\rm{P}}$.
The (mean) velocity of the motor can then be calculated as
\begin{equation}
v = l (w^+ - w^-) = k^- c_{\rm{ADP}} c_{\rm{P}} l \left [e^{\beta (\Delta \mu - \delta F l)} - e^{\beta (1-\delta) F l} \right].
\label{v}
\end{equation}

Thermodynamic quantities for each single transition can now be defined \cite{seif05, schm06a, liep08} on the basis of the transition rates. The chemical work applied during one forward step is just the chemical potential difference $\Wc = \Delta \mu$. The mechanical work delivered by the molecular motor during a single forward step against the applied force $F$ is given by $W = F l$. Since all states are equal, the internal energy does not change, $\Delta E = 0$, and thus the difference $Q \equiv  \Wc - W$ is dissipated as heat in the thermal environment.
The efficiency $\eta$ of this chemical motor is given by the ratio of mechanical work and chemical work applied by the chemical potential difference \cite{parm99} as
\beq
\eta = \frac {W}{\Wc} = \frac {F l} {\Delta \mu}.
\label{eta}
\eeq
With the force velocity relationship (\ref{v}), the power output follows as
\beq
\dot W \equiv F v = k^- c_{\rm{ADP}} c_{\rm{P}} l F \left [e^{\beta (\Delta \mu - \delta F l)} - e^{\beta (1-\delta) F l} \right].
\eeq
\begin {figure}
\begin{center}
\includegraphics[width = 0.49 \textwidth]{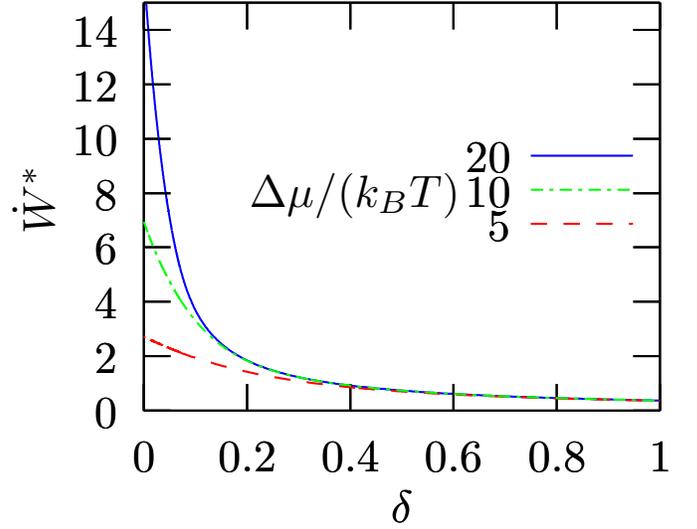}
\caption{(Color online) Maximum power $\dot W^*$ in units of $k^- c_{\rm{ADP}} c_{\rm{P}} k_{\rm{B}} T   \exp(\beta \Delta \mu)$ as a function of the position of the transition state $\delta$.
\label{fig1}}
\end{center}
\end {figure}
\begin {figure*}
\begin{center}
\includegraphics[width = 0.99 \textwidth]{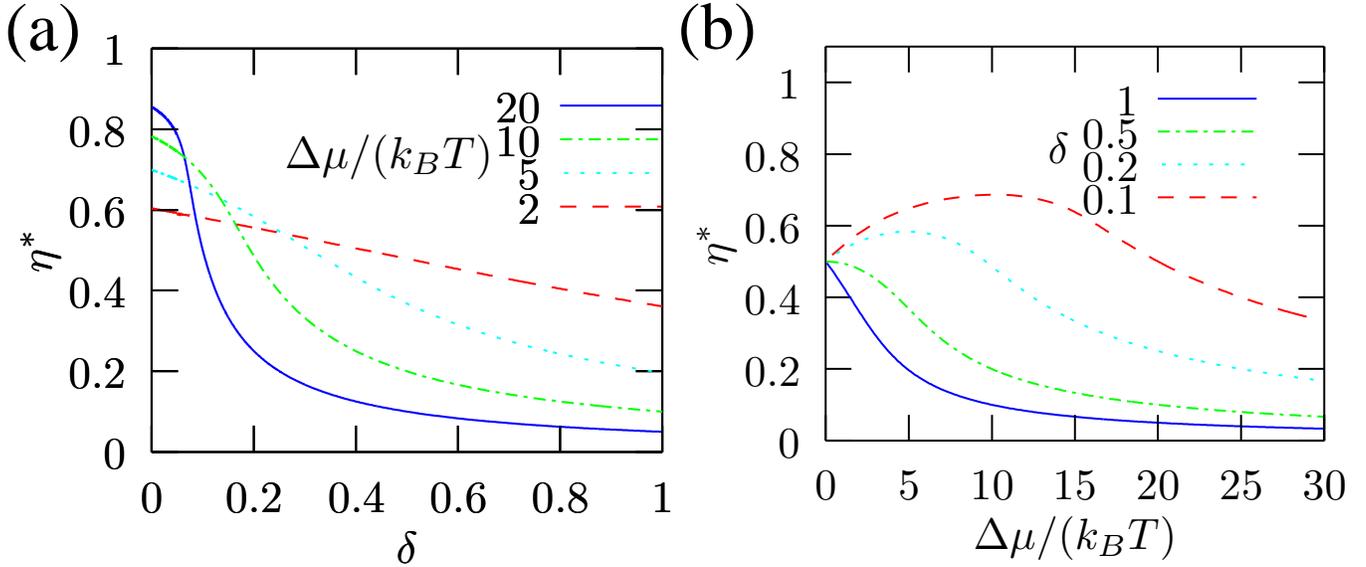}
\caption{(Color online) Efficiency at maximum power $\eta^*$ as a function of (a) the position of the transition state $\delta$ and (b) the chemical potential difference $\Delta \mu$ for Model I. 
\label{fig2}}
\end{center}
\end {figure*}
The power $\dot W$ is zero for $F \to 0$. When the force approaches the stall force $F \to \Fst \equiv \Delta \mu / l$, where the velocity vanishes, the power output also becomes infinitesimally small. Thus, there is an optimal force $\Fm$, where the power output is maximal for a given chemical potential difference $\Delta \mu$. This optimal force is given by ${d \dot W} / {d F} = 0$ which leads to the implicit relation
\beq
e^{\beta \Delta \mu } = e^{\beta l \Fm} \frac {1 +  (1-\delta) \beta l \Fm}{1- \delta \beta  l \Fm}.
\label{rel_Fm}
\eeq
For given $\beta \Delta \mu$, the scaled optimal force $\beta l \Fm$ depends only on the parameter $\delta$.   The optimal power $\dot W^*$ is shown in Fig. \ref{fig1} as a function of the transition state position $\delta$. 
Numerical results for the efficiency at maximum power $\eta^*$ are shown in Fig. \ref{fig2}. Note that the latter results are quite universal since no kinetic parameters enter these graphs. Both, power output and efficiency increase with decreasing $\delta$ and thus, a transition state near the initial position ($\delta = 0$) is most favorable. Previously, it has been speculated \cite{howa06} that such a mechanism, where forward rates are almost independent of the force, is realized in molecular motors in order to reach a large motor velocity  (corresponding to a high power output). Beyond corroborating this idea, we find as a new result that small $\delta$ also leads to a higher motor efficiency at maximum power. This is somewhat counter-intuitive since an increase in power usually leads to a decrease in efficiency.

In the limit of small chemical potential differences (where the motor works in a linear response regime near equilibrium), efficiencies at maximum power can be obtained analytically. In this limit, the stall force also becomes small and thus the exponentials in (\ref{v}) can be expanded and truncated after the first order in $\Delta \mu$ and $F$ leading to the approximate force-velocity relation
\beq
v \approx k^- c_{\rm{ADP}} c_{\rm{P}} \beta l ( \Delta \mu - F l ) .
\eeq 
In analogy to the linear response result for heat engines \cite{vdb05}, the efficiency at maximum power universally becomes $\eta^* = 1 / 2$. Beyond linear response, as a somewhat surprising result, the efficiency at maximum power increases for increasing chemical potential differences for positions of the transition state $\delta < 1 / 2$, compare Fig. \ref{fig2}b.  Usually, dissipative cost increases when the system is driven further out of equilibrium.

\section{Model II}
\begin {figure*}
\begin{center}
\includegraphics[width = 0.9 \textwidth]{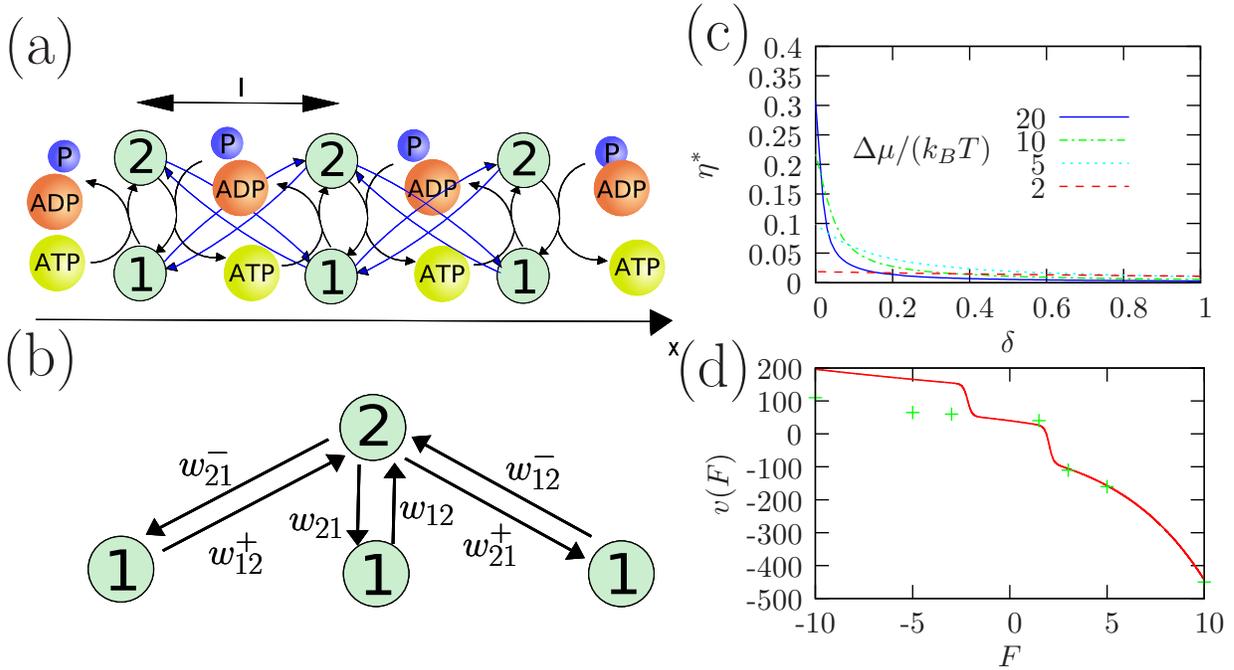}
\caption{(Color online) Two-cycle model for a molecular motor (Model II). (a) Scheme of the reaction pathways. (b) Definition of rate constants.  (c) Efficiency at maximum power as a function of the transition state position $\delta \equiv \delta_1 = \delta_2$ with $w_{21} = 0.275 / s, k_{21}^{-} = 0.7 / s, k_{21}^{+} = 1.8 / s, \Delta E = 10 k_B T$. (d) Relation between force [in $pN$] and velocity [in $nm / s$] for $\delta_1 = 0.004, \delta_2 = 0.024, w_{21} = 0.275 / s, k_{21}^{-} = 0.7 / s, k_{21}^{+} = 1.8 / s, \Delta E = 10 k_B T,  \Delta \mu = 20 k_B T,l = 9 (k_B T) / pN \simeq 36 nm$ compared to corresponding data from myosin experiments \cite{gebh06}. For these parameters, the optimal force is $F^* \simeq 1.5 pN $ and the efficiency at maximum power becomes $\eta^* \simeq 0.18$.
\label{fig3}}
\end{center}
\end {figure*}

In order to check the generality of the results obtained for the (simple) Model I, we now calculate the efficiency at maximum power for a more involved motor model. 
Recent experiments focussing on the backsteps of kinesin \cite{cart05} indicate that a realistic motor model should comprise at least one additional cycle \cite{cart05, liep07a, lau07a} leading to non-zero dissipation even at stall force. Such a mechanism with additional motor cycles presumably also applies to myosin motors which have a similar molecular structure \cite{kull96}.  In order to capture the main experimental finding of $ATP$-driven backsteps from Ref. \cite{cart05}, we propose a minimal model as shown in Fig. \ref{fig3}. 
This model is very similar to recent kinesin models \cite{cart05, liep07a, lau07a} and thus captures experimental findings qualitatively.   Binding and hydrolyzing $ATP$ leads to the unbinding of one motor head. The elastic energy then leads to a biased diffusive search for the next binding site. For high load forces, the probability of a backstep increases. Note that such backsteps involve ATP consumption and thus decrease the coupling ratio between chemical and mechanical motor cycles. The force dependence is modelled as
\begin{eqnarray}
w_{21}^+ &=& k_{21}^{+} e^{-\beta \delta_1 l F}~,~
w_{12}^+ = k_{12}^{+} e^{-\beta (1-\delta_2) l F} \nn \\
w_{21}^- &=& k_{21}^{-} e^{\beta \delta_2 l F}~,~
w_{12}^- = k_{12}^{-} e^{\beta (1-\delta_1) l F}
\end{eqnarray}
with the transition state located at $\delta_{1,2}$ for forward and backward steps, respectively.
Thermodynamic consistency requires $ k_{21}^{+} / k_{12}^{-} = \exp(\beta \Delta E)$, $k_{21}^{-} / k_{12}^{+} = \exp( \beta \Delta E)$, and $w_{12} / w_{21} = \exp[\beta (\Delta \mu - \Delta E)]$ where $\Delta E$ is the potential energy difference between state $1$ and state $2$. 
Given all rate constants, the steady state can be calculated as
\begin{eqnarray}
p_1^s &=& \frac {w_{21} + w_{21}^+ + w_{21}^-}  {w_{21} + w_{21}^+ + w_{21}^- + w_{12} + w_{12}^+ + w_{12}^- } , \\
p_2^s &=& \frac{w_{12} + w_{12}^+ + w_{12}^- }  {w_{21} + w_{21}^+ + w_{21}^- + w_{12} + w_{12}^+ + w_{12}^- }
\end{eqnarray}
with motor velocity
\beq
v = l \left[ p_1^s (w_{12}^+ - w_{12}^-) + p_2^s ( w_{21}^+ - w_{21}^-) \right ].
\eeq

The power output of the motor is $\dot W = F v$. Chemical work is applied to the motor only in the (vertical) transitions involving $ATP$ and $ADP + P$. In such a step, chemical energy of amount $\Delta \mu$ is transferred to the system and thus the chemical work per unit time is
\beq
\Pc = \Delta \mu (p_1^s w_{12} - p_2^s w_{21}).
\eeq
The motor efficiency  $\eta \equiv \dot W / \Pc$ can then be calculated for a given set of rate constants and a given force. 

We again ask for the optimal force leading to a maximal power output. We recover the qualitative results of Model I for the maximum power (data not shown) and the efficiency at maximum power, see Fig. \ref{fig3}c, also in this (more realistic) model of a molecular motor. Specifically, the largest efficiency can be achieved for both transition states near the initial position ($\delta \equiv \delta_1 = \delta_2 = 0$). For small $\delta$, the efficiency first increases with increasing chemical potential difference $\Delta \mu$ until it reaches a maximum. The advantage of a transition state near the initial position with respect to high power output and high efficiency thus seems to be a quite general characteristics for molecular motors. Note that efficiencies are generally lower due to the $ATP$-driven backward steps leading to additional dissipation in such models with additional cycles.

\section{Discussion}
In summary, we have first investigated a simple genuine model of a molecular motor under the condition of maximum power output. As our main result, we find that a transition state near the initial position yields both the largest power output and the largest efficiency at maximum power. Qualitatively, this behaviour is also recovered in a more realistic model involving a second motor cycle. 
The advantage of a small load distribution factor $\delta$ with respect to large power output and high efficiency at maximum power thus seems to be quite generic. 
We have assumed $0 \leq \delta \leq 1$ throughout this letter. Recently, it has been proposed to use $\delta < 0$ to interpret non-monotonous force velocity relations \cite{tsyg07}. For negative $\delta < 0$, the efficiency at maximum power is even larger than for $\delta = 0$ in both our model systems.

For both models, the efficiency at maximum power can increase when the system is driven further out of equilibrium by a higher chemical potential difference. This result should be distinguished from previous work predicting a maximal efficiency for non-zero chemical potential differences \cite{parm99, lau07a}. In our first model, the efficiency decreases monotonically as a function of the chemical potential difference for a given load $F$.

For kinesin motors, it is difficult to find clear evidence for a putative design principle of a transition state near the initial position. While previous studies have found small $\delta \lesssim 0.1$ for the main motor step \cite{fish01}, a transition state in the range $\delta \simeq 0.3...0.65$  has been extracted recently \cite{liep07a}. For myosin motors,
small $\delta$ has been reported for the main motor step ($\delta < 0.1$ \cite{clem05}, \cite{kolo03}) , which is also supported by our model II as detailed in the following. 

In Fig. \ref{fig3}d, the force velocity relation of our model with appropriate rate constants is compared to a recent myosin experiment \cite{gebh06}. For forces $F>0$, the experimental data, including a step at $F \simeq \Delta \mu / l$, is captured by our simple model.  For negative forces, our model shows a second step at $F \simeq -\Delta \mu / l$ which is not present in the experimetal data. Both steps in the theoretical model are explained by the fact that for large forces $|F| \gg \Delta \mu / l$, vertical (ATP driven) transitions are slow compared to the  horizontal (force dependent) transitions and the motor is basically pulled by the load. The step at negative forces could be eliminated by introducing a force dependent energy difference $\Delta E(F)$ which corresponds to an additional force dependence of the vertical transitions. However, instead of introducing new parameters, we keep the simple model II and extract from the crude fit shown in Fig. \ref{fig3}d transition state positions $\delta_1 = 0.004$, $\delta_2 = 0.024$. These values mainly determine the behaviour of the molecular motor at large loads $| F | \gg \Delta \mu / l$. With these parameters, we find an efficiency at maximum power of $\eta^* \simeq 0.18$ which, given the strong decay of $\eta^*$ with increasing $\delta$, is quite close to the optimal value $\eta^* \simeq 0.31$ for $\delta_1 = \delta_2 = 0$, see Fig. \ref{fig3}c.

We do not claim that our simple model can explain all aspects of myosin motility. Rather, we have chosen model II in order to probe the robustness of our main results concerning the efficiency at maximum power. In order to construct a comprehensive model of myosin, more experimental data seem to be necessary. 

If future experiments confirm the indication that myosin motors have a transition state near the initial position for the main motor step (corresponding to an almost force-independent forward rate), it would be tempting to speculate whether evolutionary pressure for efficiency and large power has led to this characteristic. For a more comprehensive answer to this question, however, other evolutionary goals like speed, robustness,  and high processivity should be considered. Likewise, the dependence of our results on the interaction between single motor domains needs to be explored in future work.

\end{document}